\documentclass[12pt]{article}
\usepackage{amssymb,amsmath,amscd}
\begin{document}

\begin{titlepage}

                            \begin{center}
                            \vspace*{0cm}
\Large\bf{Tsallis entropy induced metrics and CAT(k) spaces}\\

                            \vfill

              \normalsize\sf    NIKOS \  \ KALOGEROPOULOS $^\star$\\

                            \vspace{0.2cm}
                            
 \normalsize\sf Weill Cornell Medical College in Qatar\\
 Education City,  P.O.  Box 24144\\
 Doha, Qatar\\

                            \end{center}

                            \vfill

                     \centerline{\normalsize\bf Abstract}
                     
                           \vspace{3mm}
                     
\normalsize\rm\setlength{\baselineskip}{18pt} 

\noindent Generalizing the group structure of the Euclidean space,  we construct a Riemannian metric 
on the deformed set \ $\mathbb{R}^n_q$ \ induced by the Tsallis entropy composition property. 
We show that the Tsallis entropy is a  ``hyperbolic analogue" of the ``Euclidean" 
Boltzmann/Gibbs/Shannon entropy and find a geometric interpretation for the nonextensive parameter $q$. 
We provide a geometric  explanation of the uniqueness of the Tsallis entropy as reflected through 
its composition property, which is provided by the Abe and the Santos axioms. 
For two, or more, interacting systems described by the Tsallis entropy, having different values
of  $q$,  we argue why a suitable extension of this construction is provided by the 
Cartan/Alexandrov/Toponogov metric spaces with a uniform negative curvature upper bound. \\    

                             \vfill


                             \vfill

\noindent\rule{8cm}{0.2mm}\\
   \noindent \small\rm $^\star$  E-mail: \ \  \small\rm nik2011@qatar-med.cornell.edu\\

\end{titlepage}


                                                                                        \newpage

                                                          \normalsize\rm\setlength{\baselineskip}{18pt}

                                                                  \centerline{\large\sc 1. \ Introduction}

                                                                                       \vspace{5mm}

The Tsallis entropy was introduced almost two decades ago [1], and  since its inception it has attracted 
considerable interest in the Physics community as well as outside it [2] for a variety of reasons. Among these reasons, one
could include its wide conjectured applicability in describing collective phenomena with long-range spatial and temporal 
correlations. Related to these, are  systems exhibiting a fractal evolution on their phase space [2] (and references therein). 
Initial indications seem to support some of these claims, mostly by using results of numerical simulations of relatively 
simple systems (``toy models") [2]. However, the scope of such results, their domain of applicability and precise 
model-independent conclusions  are still subject to investigation [3].\\   

A derivation of the Tsallis entropy from the microscopic dynamics of a dynamical system is a highly desirable goal [2]. This is part of Boltzmann's 
approach on Statistical Mechanics. Such a derivation has not yet been completed yet, even for the far better known 
Boltzmann/Gibbs/Shannon entropy (henceforth abbreviated as BGS entropy), despite the substantial progress made in the latter case. 
The corresponding problem for the Tsallis entropy is more difficult, as there are still 
formal aspects of it that need to be better understood. Pertinent for our purposes is that one such problem is associated 
with the composition property of the Tsallis entropy. This composition (``generalized addition") has forced us to re-evaluate the meaning of the 
concepts of ``independence" and ``additivity",  in particular for systems with long-range interactions that the Tsallis 
entropy conjecturally describes [2]. Attempts to define a generalized multiplication which is distributive with respect to the generalized 
addition and which would also  reflect the composition properties of the Tsallis entropy have only very recently come to fruition [4], [5].\\

In [5], we independently re-discovered such a generalized multiplication which had been previously constructed by [4]. 
Although the explicit equivalence between the multiplications introduced in [4], [5] is still lacking, the simplicity and similarity of  the
final results makes such an equivalence almost inevitable.
To achieve our goals,  we defined in [5], as was also previously done in [4] although not in this notation, a deformation of the reals indicated by 
\ $\mathbb{R}_q$ \ and a field isomorphism \ $\tau_q : \mathbb{R} \rightarrow \mathbb{R}_q$. \  We explored some of the 
induced metric- and measure-theoretical aspects of \ $\tau_q$ \ for the set \ $\mathbb{R}_q$ [5]. \ In the present work we 
continue the investigation along the same lines, focusing on constructing metrics induced by \ $\tau_q$, \  on \ $\mathbb{R}^n$ \ in an attempt to 
get a firmer grasp of the properties of \ $\tau_q$ \ to more realistic cases. Exploring such properties can provide a better 
understanding of the composition properties of the Tsallis entropy hence of the concepts of independence and additivity that give rise to it.  
In Section 2, we construct two metrics induced by the Tsallis entropy. The construction of the Riemannian
metric relies on the group-theoretical properties of \ $\mathbb{R}^2$ \ induced by the Tsallis entropy composition. 
In Section 3, by using results on warped products, we find that this metric has constant negative sectional curvature.  
Then, we provide an interpretation of the entropic parameter 
$q$ in terms of this curvature and comment on the  uniqueness of the Tsallis entropy encoded in its composition property, 
from our geometric perspective. In Section 4, we extend the formalism to interacting systems with different values of the entropic parameter $q$, 
superstatistics being a notable example, and propose that a general and flexible metric framework reflecting the composition properties of the
Tsallis entropy is provided by the category of CAT(k) spaces.  Section 5 briefly alludes to preliminary results which generalize and extend 
the current work.\\ 

                                                                                           \vspace{5mm}


                                                  \centerline{\large\sc 2. \  Metrics induced by the Tsallis entropy}

                                                                                            \vspace{5mm}

The Tsallis entropy is a one-parameter family of entropic functionals \ $S_q$, \ which are parametrized by \ $q\in\mathbb{R}$, \ called the 
non-extensive or entropic parameter.  To define the Tsallis entropy, consider a discrete set of outcomes (sample space)  whose elements 
are parametrized by \ $i\in I$ \ with \ $I$ \ being an appropriate discrete set. Let each outcome occur with probability \ $p_i$. \ 
Then the Tsallis entropy was defined [1] by 
\begin{equation} 
    S_q = k_B \  \frac{1}{q-1} \left( 1 - \sum_{i\in I} \ p_i^q \right)
\end{equation}
If the sample space is continuous \ $\Omega$, \ then the set of outcomes is characterized by a probability density function $p(x), \ x \in \Omega$ \ and the 
Tsallis entropy is defined by the continuous analogue of (1), namely 
\begin{equation}
    S_q = k_B \  \frac{1}{q-1} \left( 1 - \int_{\Omega} \left[ p(x) \right]^q dx \right)
\end{equation} 
where \ $dx$ \ is an appropriate integration measure, the Hausdorff measure of \ $\Omega$ \ being the most common choice, and \ $k_B$ \ is Boltzmann's
constant, henceforth to be set \ $k_B = 1$ \ for simplicity. It is straightforward to check that the Tsallis entropy reduces to the BGS entropy when \
$q\rightarrow 1$, \ namely
\begin{equation}
   \lim_{q\rightarrow 1} \ S_q = S_{BGS}
\end{equation}
It can also be immediately checked that if two sets of outcomes \ $\Omega_1$, $\Omega_2$ \ are independent, in the conventional sense of the word, 
namely if they satisfy 
\begin{equation}
   p_{\Omega_1 + \Omega_2} = p_{\Omega_1} \cdot p_{\Omega_2}   
\end{equation}
then the corresponding Tsallis entropies obey the generalized composition property
\begin{equation}
   S_q (\Omega_1 + \Omega_2) = S_q (\Omega_1) + S_q (\Omega_2) + (1-q)  S_q(\Omega_1) S_q(\Omega_2) 
\end{equation}
Expectedly, this recovers the usual additivity of the BGS entropy for independent outcomes, in the limit \ $q\rightarrow 1$, \ as can also be readily seen. 
The composition property of the Tsallis entropy (5) has forced us to re-consider the definitions of independence and additivity [2]. The latter was 
generalized [6], [7] to mirror (5), by
\begin{equation} 
   x \oplus_q y = x + y + (1-q) x y
\end{equation}
What was lacking for some time was a generalized multiplication that would be induced by the  Tsallis entropy, which would be the counterpart to (6). 
Then (5) and (6) together would form an algebraic structure, such as ring or a field that would allow further mathematical structures of potential 
physical interest to be defined in it.  Proposals for such a generalized product gradually emerged [6]-[9]
but they seemed to have some other undesirable characteristics, hence the problem remained open for a while [3]. The difficulty could be traced on how to 
define a generalized product induced by the Tsallis entropy that was ``natural" enough, be able to recover the ordinary product in the limit 
\ $q\rightarrow 1$ \  and at the same time make that product distributive with the generalized sum (6). The problem was solved in [4], 
and later independently in [5], when it was realized that a way to define such a generalized multiplication would be accomplished by deforming 
not only the operation of the multiplication itself but also the underlying set.  
To formalize this, in [4], [5] the deformation of the set of reals, indicated by  \ $\mathbb{R}_q$, \ was defined via the corresponding field 
isomorphism
 \ $\tau_q: \mathbb{R}\rightarrow \mathbb{R}_q$ \ 
\begin{equation}
   \tau_q (x) = \frac{(2-q)^x - 1}{1-q} \equiv x_q
\end{equation}
in the notation of [5]. Although the definition of the generalized product in [4] indicated by \ $\Diamond_q$, \ and in [5] indicated as \ $\otimes_q$, \ look 
different, simplicity and identity of results encoded in (7)  indicate that they should be equivalent. 
Notice that this equivalence is indirect,  as a desirable explicit  isomorphism between these generalized products is still lacking. 
In the present work, we will be using the form of the generalized multiplication   
provided in [4], rather than in [5], as it is in closed form, so it is easier to work with for our purposes. 
This generalized product \ $\Diamond_q$ \ was introduced in eq. (24) of [4] as
\begin{equation}
    x_q \Diamond_q \ y_q = \frac{(2-q)^{\frac{\log [1+(1-q)x] \cdot \log [1+(1-q)y]}{[\log (2-q)]^2}}  -1}{1-q}
\end{equation}
and can be re-expressed in the notation of [5] via \ $\tau_q$  
\begin{equation}
       \tau_q (x\cdot y) = x_q \Diamond_q \ y_q = \tau_q \left( \tau^{-1}_q (x_q)\cdot  \tau^{-1}_q (y_q) \right)
\end{equation}
where \ $\cdot$ \ stands for the usual multiplication in \ $\mathbb{R}$, \ which is easily understood and will be omitted henceforth. 
These multiplications  make the following diagram commutative 
   \begin{equation}
       \begin{CD}
         \mathbb{R} \times \mathbb{R}                            @ > \cdot >>                                  \mathbb{R} \\
                    @V \tau_q \times \tau_q VV                                                                              @VV \tau_q V\\
         \mathbb{R}_q\times\mathbb{R}_q        @ >> \otimes_q \ , \ \Diamond_q >         \mathbb{R}_q        
        \end{CD} 
   \end{equation}
In the present work our aim is to examine some of the metric implications of (7),  hence of the Tsallis entropy,
generalizing and extending results of [5]. \\

The form of \ $\tau_q$ \ in (7)  leads us to suspect a ``natural" distance function that could be defined on \ $\mathbb{R}^n$ \ to explore 
\ $\tau_q$'s metric consequences.  To do so, we observe that by defining the evaluation map ``absolute value" 
\ $| \cdot |_q$ in \ $\mathbb{R}_q$ \ by demanding
\begin{equation}
|x-y| = |x_q \ominus_q y_q|_q 
\end{equation}
we get an isometry between \ $\mathbb{R}$ \ and \ $\mathbb{R}_q$,  \ by definition. This is not very enlightening though, if we want 
to compare the metric properties of \ $\mathbb{R}$ \ and \ $\mathbb{R}_q$. \ 
An analogy with Special Relativity may be of some interest at clarifying the situation and our goal. 
Consider two inertial observers moving with a relative speed \ $v$. \ Let these two observers compare 
the lengths of two sticks of equal proper length. Each observer gets one of these sticks and places it along the direction of relative motion. 
By definition, each observer, measures the length of its own stick to be its proper length. The non-trivial characteristics of space-time emerge 
when each observer compares the length of its stick with that of the stick of the other observer as both are measured by him/her. 
Each observer will get a different answer about the length of its stick as compared to the length of the stick of
the other observer. Such measurements are compared by using the Lorentz transformations (length contraction) 
between the two inertial frames, which are parametrized by $v$. In turn, 
these Lorentz transformations include a lot of information about the Minkowski metric of space-time, as they are generators of its isometry 
group.  Getting back to the case of the Tsallis entropy, the analogue of the Lorentz transformations of the relativistic example is provided by the 
field isomorphism \ $\tau_q$. \ This isomorphism should induce a metric allowing the comparison of distances between \ $\mathbb{R}$ \ and
 \ $\mathbb{R}_q$ \ as are both measured from the viewpoint of the same set, be it either \ $\mathbb{R}$ \ or \  $\mathbb{R}_q$. \ 
Using the above analogy, the nonextensive parameter  \ $q$ \ plays a role similar to the relative speed \ $v$ \ of the relativistic example. 
The BGS entropy, being the limit of the Tsallis entropy (3) , \ corresponds to the ``Newtonian"/non-relativistic limit \ $v\rightarrow 0$. \ 
Hence, the induced metric structure 
by the Tsallis entropy composition property (5), should reduce to that induced by the BGS entropy for \ $q\rightarrow 1$, \ in the same way as the Lorentz 
transformations reduce to the Galilean transformations in the non-relativistic limit. What we are actually interested in, is the converse: a generalization of 
the metric structure induced by the ``simple" addition of the BGS entropy to that of the Tsallis entropy through \ $\tau_q$. \ It should be stressed, that the 
above is just a motivating analogy. No claim is made about any deep common physical origin or connection between the induced metric structure by the 
Tsallis entropy and the Minkowski metric of  Special Relativity.\\
 
Motivated by this analogy and by (7), (11), a reasonable choice for a distance function 
\ $\tilde{d}: \mathbb{R} \times \mathbb{R} \rightarrow \mathbb{R}$ \ would be \ $\tilde{d} = \tau_q(d)$ \ or, explicitly,
\begin{equation}
\tilde{d}(x,y) = \frac{(2-q)^{d(x,y)}-1}{1-q}
\end{equation}  
where we have set \ $d(x,y) = |x-y|$ \ The definition can be carried over to the case of an induced distance function \ $\tilde{d}$ \ in \ $\mathbb{R}^n$ \ 
which can also be defined by (12). This can be further  generalized  by declaring \ $\tilde{d}$ \  to be the distance function  induced by \ $\tau_q$ \ 
irrespective of whether \ $d(x,y)$ \ is induced by the absolute value or its linear space generalization, the \ $l^1$ \  Banach norm, as in (11), or not.\\ 

One can straightforwardly verify that \ $\tilde{d}(x,y)$ \ is indeed a distance function. It is obviously symmetric in its arguments
\begin{equation}
  \tilde{d}(x,y) = \tilde{d}(y,x), \hspace{5mm} \forall \ x, y \in X
\end{equation}  
where \ $(X, d)$ \ stands for a general metric space, although we will almost exclusively be using \ $X=\mathbb{R}^n$ \ in the sequel. 
Moreover, 
\begin{equation}
   \tilde{d}(x,y) = 0  \hspace{5mm} \iff \hspace{5mm} x=y
\end{equation}
and since \ $d(x,y)\geq 0$ \ and \ $q \in [0,1)$, \  then \ $\tilde{d}(x,y)$ \ is  
positive-definite. To check that \ $\tilde{d}(x,y)$ \ indeed satisfies the triangle inequality, we assume that \ $d(x,y)$ \ does so, namely that
\begin{equation}
  d(x,y) + d(y,z) \geq d(x,z), \hspace{5mm} \forall \ x,y,z \in X
\end{equation}
 Then
\begin{eqnarray}
   \tilde{d}(x,y) + \tilde{d}(y,z)   =  & \frac{(2-q)^{d(x,y)}-1}{1-q} + \frac{(2-q)^{d(y,z)}-1}{1-q}\\
                                                     =  &  \frac{2}{1-q}  \{  \frac{(2-q)^{d(x,y)} + (2-q)^{d(y,z)}}{2}  -1   \}  
\end{eqnarray}
Using the arithmetic-geometric mean inequality
\begin{equation}
     \frac{a+b}{2} \geq \sqrt{ab}
\end{equation}
we obtain
\begin{equation}
 \tilde{d}(x,y) + \tilde{d}(y,z) \geq \frac{2}{1-q} \{ (2-q)^{\frac{d(x,y)+d(y,z)}{2}} -1 \}
\end{equation}
which due to (12) gives
\begin{equation}
\tilde{d}(x,y) + \tilde{d}(y,z) \geq 2 \cdot \frac{(2-q)^{\frac{d(x,z)}{2}} - 1}{1-q}
\end{equation} 
which by using (9) amounts to
\begin{equation}
   \tilde{d}(x,y) + \tilde{d}(y,z) \geq \tilde{d}(x,z)
\end{equation}
as is required.\\

The inconvenient aspect of the distance function \ $\tilde{d}$ \ is that it is not associated to a Riemannian 
metric, even if \ $d$ \  is. Since, we want to remain within the class of Riemannian metrics we proceed as follows:
 Riemannian metrics are up to first order approximation Euclidean.
Differences between flat and curved spaces appear as second order approximations, where the Riemann tensor 
is introduced as a multilinear object, whose components quantify such differences [10]. However, Euclidean spaces have an
extra structure: they are groups. Or to be more specific, they are Abelian groups under translations. Given this basic group structure
of \ $\mathbb{R}^n$, \ a way to construct a Riemannian metric which is not  flat to second order, or equivalently a way to 
make the space curved, is by changing the composition properties of the translations, or equivalently by modifying or ``twisting"
the definition of translations. Such ``twists" are quite common in Algebra, and  have  been extensively used 
in group theoretical applications in Physics [11]. Since we want to compare the properties 
of \ $\mathbb{R}$ \ and \ $\mathbb{R}_q$, \ the simplest set on which to realize such a comparison is the Cartesian product 
\begin{equation}        
   (x,y) \ \in \  \mathbb{R} \times \mathbb{R} 
\end{equation}
We choose, the first entry to follow the composition of the standard translations, namely to behave additively. We use the standard 
addition/composition of the first entry as a ``reference" against which the generalized addition of the second entry will be contrasted.  
To introduce the ``twist", we compose the second entry
in a way that reflects the essential difference between \ $\mathbb{R}$ \ and \ $\mathbb{R}_q$. \ Such a difference is expressed
by (7)  which essentially calculates the exponential, with basis \ $2-q$ \ although this is of little importance, 
of an element of \ $\mathbb{R}$, \ followed by a constant 
shift. Since a constant shift appears to be of little interest, at a first glance, we will focus on the exponential action of \ $\tau_q$. \ 
To wit, we define the composition (``multiplication") of the  Cartesian pairs to be given by the semi-direct product 
\begin{equation}
   (x,y) \ltimes  (x', y') = (x+x', y+e^{tx}y')
\end{equation}
We have introduced the parameter \ $t \in \mathbb{R}$  \ in order to allow some flexibility in the proposed structure. 
It is straightforward to check that  \ $\ltimes$ \ thus defined on \ $\mathbb{R}^2$ \ is associative, has identity element
\ $(0,0)$ \ and inverse element
\begin{equation}
  (x,y)^{-1} = (-x, -e^{-tx}y)
\end{equation}    
Hence, the resulting structure maintains the essentials of the group structure of \ $\mathbb{R}^2$ \ under translations, 
but the effect of the ``twist" of \ $\ltimes$ \ makes it manifestly non-Abelian. However such a twist is not too ``exotic", 
so the resulting group structure \ $G = (\mathbb{R}^2, \ltimes )$ \ remains ``almost" Abelian in the following sense. 
Consider the first-step commutator subgroup \ $G^{(1)} = [G, G]$, \ which is generated by all elements of \ $G$ \ of the form   
\begin{equation}
    G^{(1)} = \langle \ (x_1, y_1) \ltimes (x_2, y_2) \ltimes (x_1, y_1)^{-1} \ltimes (x_2, y_2)^{-1}, \  \  \
                                                x_i, y_i \in \mathbb{R}, \ \ \ i = 1, 2 \ \rangle
\end{equation}
Iterating this, consider the second-step commutator subgroup by setting \ $G^{(2)} = [G^{(1)}, G^{(1)}]$, \ and inductively
\ $G^{(n)} = [G^{(n-1)}, G^{(n-1)}], \ \ n \in \mathbb{N}$. \ We see, by a direct calculation, that in the case of \ $(G, \ltimes)$ \ 
this iteration of  commutator subgroups stops at the second step, meaning that \ $G^{(2)}$ \ is the identity (trivial) element.  
It is in this sense that $(G, \ltimes)$ is ``almost" Abelian, since for an  Abelian group, by definition, \ $G^{(1)}$ \ is trivial.\\ 

We can also interpret the semi-direct product (23), as the action of \ $G\times G \rightarrow G$ \ by left translations
\begin{equation} 
    L_{(x,y)} (x',y') = (x,y) \ltimes (x',y')
\end{equation}
Considering that  \ $G=\mathbb{R}^2$, \ is not just a group but it has also a compatible differentiable structure, 
being a Lie group,  we want to construct a left-invariant Riemannian metric on it, \ given by a metric tensor \ $\bf{g}$ \ on which 
the left translations act as isometries. We do this in order to
keep the similarity of the proposed structure with the homogeneity of the Euclidean spaces. 
Consider  the ``abscissa" \ $(x,0), \ x \in \mathbb{R}$. \  A left translation shifts it to 
\begin{equation}   
    L_{(x_0, y_0)} (x, 0) = (x_0+x, y_0)
\end{equation}
Similarly, the ``ordinate" \ $(0,y), \ y \in \mathbb{R}$ \ is mapped to  
\begin{equation}
    L_{(x_0, y_0)} (0, y) = (x_0, y_0+e^{tx_0}y)
\end{equation}
The Jacobian matrix \ $J_{(x_0,y_0)}$ \ of this left translation is
\begin{equation}
   J_{(x_0,y_0)} =   \left(                  
                    \begin{array}{cc}
                                        1 & 0\\
                                        0 & e^{tx_0}
                    \end{array}  \right)
\end{equation}
So, the tangent vectors to the abscissa \ $(dx, 0)$ \ and ordinate \ $(0, dy)$ \ get shifted under \ $dL_{(x_0,y_0)}$ \ to  
\begin{equation}
 \left( \begin{array}{c}
               dx'\\
               dy'
          \end{array}\right)  = J_{(x_0,y_0)} \left( \begin{array}{c}
                                                                                     dx\\
                                                                                     dy
                                                                               \end{array} \right)
\end{equation}
giving 
\begin{equation}
     (dx', 0) = (dx, 0)
\end{equation}
and
\begin{equation}
     (0, dy') = (0, e^{tx_0}dy)
\end{equation}
These two vector fields are orthonormal at the origin of \ $G$ \ with respect to the usual Euclidean metric \ ${\bf g_E}$. \ Using (32), 
the only non-trivial condition that  has to be satisfied to determine the required left-invariant metric \ ${\bf g}$ \ under translations is
\begin{equation}
   {\bf g_E}(dy, dy)|_{(0,0)} = {\bf g_E}(e^{tx_0}dy, e^{tx_0}dy )|_{(x_0,y_0)}
\end{equation}
which due to the bi-linearity of \ ${\bf g_E}$ \ amounts to 
\begin{equation}   
    {\bf g_E}(dy, dy)|_{(x_0,y_0)} = e^{-2tx_0} {\bf g_E}(dy, dy)|_{(0,0)}
\end{equation}
Following this, we define the left-invariant under translations, metric \ $\bf{g}$, \ with components arranged in matrix form, by
\begin{equation}
     \bf{g} = \left( \begin{array}{cc}
                               1 & 0 \\
                               0 & e^{-2tx}
                           \end{array} \right)     
\end{equation}
with the corresponding infinitesimal line element  given by
\begin{equation}
   ds^2 = dx^2 + e^{-2tx}dy^2
\end{equation}
The two equations (350, (36) are the sought after metric which is induced by (5).
We see that the lines $y$: const. are geodesics, namely the embeddings
\ $\mathbb{R}\rightarrow (\mathbb{R}, y)$ \ are isometric. By contrast, the lines $x$:const. are curved, 
and shrink exponentially fast as $x$ increases.
Moreover  we could have even be able to guess from the outset the form of (35). Indeed, 
it is  the diagonal metric given by $(1, e^{-2tx})$. The first entry reflects that the first factor in $G$ is unaffected by 
the ``twist" (23), so it is a simple translation. The second factor is essentially \ $\tau_q^{-1}$, \ if one first omits the 
constant shift due of \ $-1/(1-q)$ \ in \ $\tau_q$ \ and after switching the base of the exponentials 
\ $2-q$ \ with \ $e$, \ as noted above. 
A point of interest, and potential problems, arises when we observe that the constant shift by \ $-1/(1-q)$ \ may have graver 
consequences than initially thought. This occurs because the naive guess 
\begin{equation}       
   ds^2 = dx^2 + \tau_q(x) ^{-2} dy^2
\end{equation}
becomes degenerate when \ $x=0$. \ One way to avoid problems associated with this degeneracy is 
to focus on a particular subset of \ $\mathbb{R}^2$, \ namely on the points that are far away from each other 
and from the origin. In essence, this amounts to confining our arguments on  the subset of \ $\mathbb{R}^2$ \ whose elements are  the 
``points at infinity", which have to be properly defined. It turns out that using this restriction is sufficient for drawing the conclusions 
of interest for this work.\\

It should be pointed out  that the construction of the above Riemannian metric is a very special case of the 
vastly more general and elaborate constructions on homogenous spaces of non-positive curvature 
via solvable groups with left-invariant metrics [12],  [13],  [14].\\   

                                                                      \vspace{5mm}


                                                                    \newpage

       \centerline{\large\sc 3. \   Negative \ Curvature \ and  \ Tsallis \ Entropy \ Universality }

                                                                       \vspace{5mm}

As is well-known, the most important local quantity related to the metric on a Riemannian manifold is the Riemann tensor [10].
Previously we were dealing with metric on \ $\mathbb{R}^2$, \ for which the only non-trivial quantity 
derived from the Riemann tensor is the Gaussian curvature \ $k$. \ The metric tensor (35) is diagonal, so a straightforward 
computation gives \ $k$. \ In view of the arguments and generalizations 
proposed below in this Section, we choose to take a different path. We observe that (35) is a warped product  [15]. 
To define warped products, consider the  product manifold \ $M = B\times F$ \ 
where both \ $B$ \ and \ $F$ \ are Riemannian manifolds with corresponding metric tensors \ ${\bf g}_B$ \ and \ ${\bf g}_F$ \  and where \ 
$\times$ \ denotes the topological product.  Let \ $p_1: M\rightarrow B$ \ and \ $p_2: M\rightarrow F$ \ be projections on the 
first ($B$: ``the base") and second ($F$: ``the fiber") factors of the topological product, respectively. 
Warped product metrics \ ${\bf g}_M$ \ on 
\begin{equation}
   M = B \times_f F 
\end{equation}
generalize the Riemannian product metrics on \ $B\times F$, \ by 
``twisting" them by a function \ $f: M\rightarrow \mathbb{R}$ \ as follows
\begin{equation}     
   {\bf g}_M = {\bf g}_B + f^2(b) {\bf g}_F
\end{equation}
where \ $b\in B \subset M$. \ Probably the simplest case of a metric being in a warped form
is the Euclidean metric on \ $\mathbb{R}^n$ \ expressed in polar coordinates. 
As can be immediately seen, the warped product metrics are generalizations of the 
induced metrics on surfaces of revolution in \ $\mathbb{R}^n$. \ 
Due to their highly symmetric properties, warped product metrics are 
very frequently encountered in Physics, most notably in theories of gravity such as General Relativity [16] or 
String/M-theory models [17]. Warped product metrics can also be seen as  special cases of Riemannian submersions [18] 
with horizontal leaves \  $p_2^{-1}(x), \  x \in F$ \ which are totally geodesic 
submanifolds of \ $M$. \  The restriction of \ $p_1$ \ to \ $p_2^{-1}(F)\rightarrow B$ \ is an isometry.
Hence the corresponding second fundamental form (shape operator) of the horizontal leaves of the foliation 
(the \ $A$ \ tensor in the notation of [18]) vanishes identically. In our case, both \ $B$ \ and \ $F$ \ are isometric to \ 
$\mathbb{R}$ \ and the warping function of interest is the one-parameter family of exponentials 
\begin{equation}
    f_t(x) = e^{-2tx}
\end{equation}
which simplifies considerably the expressions derived in [18], [15] for the sectional curvature. These simplify to  
\begin{equation}  
     k = - \frac{f_t''(x)}{f_t(x)}
\end{equation}    
which finally gives  
\begin{equation}
       k = -t^2
\end{equation}
To go back the special case of interest for the Tsallis entropy, all that we have to do is to set 
 \begin{equation}
    t = \log (2-q)
 \end{equation}
In such a case, the  metric tensor (35) is of  constant negative curvature
\begin{equation}   
      k = - [ \log (2-q)]^2 
\end{equation}
We observe that what the semi-direct product (23)  accomplished, was to provide a map from \ $\mathbb{R}^2$ \ equipped with
the Euclidean metric to \ $\mathbb{R}^2$ \ equipped with the constant negative curvature metric (35). The definition of independence 
reflected via the composition of the BGS entropy is through the usual addition. 
The corresponding group structure on \ $\mathbb{R}^2$ \  that geometrically reflects this fact is that of the translations. 
Hence the corresponding left-invariant metric is the usual $l^2$ ``Euclidean" one. 
By contrast, the generalized addition (6) defines a new type of independence for the Tsalis entropy.  This induces the semi-direct product structure (23)
which results in the left-invariant negative curvature metric (35) on \ $\mathbb{R}^2$. \ 
Because of this,  the two definitions of independence, which stem from the difference in the composition properties of the corresponding entropies (6) for 
\ $q\neq 0$ \ and \ $q=0$, \ allow us to state that the Tsallis entropy is a ``hyperbolic counterpart" or a ``hyperbolic analogue" of the ``Euclidean" 
BGS entropy.  Predictably, as  \ $q\rightarrow 1$, \ the curvature (44) of the hyperbolic metric (35), (43) becomes zero, thus recovering
the BGS ``flat" case. \\

The previous discussion gives rise to a geometric interpretation of the generalized multiplication (8) in view of the curvature result (44). 
The numerator of the exponential of (8) is the area of a Euclidean rectangle whose sides have lengths  \  $\tau_q^{-1}(x_q)$ \  and  \  $\tau_q^{-1}(y_q)$. \ 
The denominator of the exponential of (8) is the absolute value of the curvature \ $k$ \ given in (44). It may be worth re-interpreting this exponent in 
light of the following, equivalent, definition of the sectional curvature \ $k$, \  known as the Bertrand-Diquet-Puiseux theorem:  
Let \ $x\in M$ \ be a point on the Riemannian manifold \ $M$, \ and let \ $U, V \in T_xM$ \ and \
 $\Sigma = U \wedge V$ indicate the linear subspace of \ $T_xM$ \ spanned by \ $U$ \ and \ $V$. \ Let \ $D(r)$ \ be the 2-dimensional disk of 
 radius \ $r$ \ in \ $\Sigma$, \ 
centered at $x$  \  and let the image of \ $D(r)$ \ under the exponential map have area \ $A(r)$. \ Since the area of \ $D(r)$ \ is \ $\pi r^2$, \ then       
\begin{equation}
      k_\Sigma (x) = \lim_{r \rightarrow 0} \ \frac{12}{r^2}  \left(1 -  \frac{A(r)}{\pi r^2}\right) 
\end{equation}
In words, the sectional curvature \ $k_\Sigma (x)$ \ determines the relative rate of increase, or decrease, of the area of a small disk lying on \ $M$, \
centered at \ $x \in M$ \ which is generated by the radial geodesics, with respect to its Euclidean counterpart of the same radius lying on \ $\Sigma$. \  
From a comparison of (8), (44) and (45), we can see that the exponent of (8) gives, almost, the area of a quadrilateral of side 
lengths \ $\tau_q^{-1}(x)$ \ 
and \ $\tau_q^{-1}(y)$ \ in \ $(\mathbb{R}^2, {\bf g})$. \ Then, the generalized multiplication (8) gives the area of this rectangle in \ $\mathbb{R}_q^2$ \ 
as measured by its own Euclidean (flat) metric, a fact which is in accordance with the geometric meaning of the ordinary multiplication 
in \ $\mathbb{R}^2$ \ as also becomes evident in (9).  \\  

 The expression (44) for \ $k$ \  provides another geometric  interpretation for the entropic parameter: \ $q$ \  determines the 
curvature \ $k$ \ of the metric through (44). Since the curvature uniquely determines the metric up to isometry for simply-connected constant curvature 
manifolds, then \ $q$ \ also determines uniquely \  $\bf{g}$. \ This is a somewhat convoluted way of expressing  
a fact which is evident by combining (35) and (43).  As a result, \ $q$ \ determines all the metrically-related features of  \ $(G=\mathbb{R}^2, \bf{g})$. \\ 

The above group theoretical construction can be generalized to higher dimensions in a very straightforward  manner. Instead of considering 
Cartesian pairs, we could have considered Cartesian $(n+1)$-plets  \ $\{ (x_0, x_1, \ldots, x_n) \in \mathbb{R}\times\mathbb{R}^n, n\geq2) \} $ \ 
where the semi-direct product \ $\ltimes$ \ ``twists", using the zeroth entry,  all subsequent entries except the zeroth entry itself, namely
\begin{equation}
  (x_0, \ x_1, \ \ldots, \ x_n) \ltimes (x_0', \  x_1',\  \ldots, \ x_n') = (x_0 + x_0', \ x_1+ e^{tx_0}x_1', \  \ldots, \ x_n + e^{tx_0}x_n')  
\end{equation}
This results in endowing the space of interest \  $G = (\mathbb{R} \times \mathbb{R}^n, \ltimes )$ \
 with the diagonal left-invariant Riemannian metric \ ${\bf g} = (1, e^{-2tx_0}, \ldots, e^{-2tx_0})$ \ having  
infinitesimal line element given by
 \begin{equation}
     ds^2 =  dx_0^2 + e^{-2tx_0} (dx_1^2 + \ldots + dx_n^2 )
\end{equation}
and \ ${\bf g}$ \ having constant negative sectional curvature \ $k=-t^2$ \ as in (42).\\

We saw that the Tsallis entropy composition (5), reflecting a generalization of the concept of independence,
induces the left-invariant constant negative curvature (42)
distance function (47) on \ $\mathbb{R}^n$. \  As a result, the metric properties induced by the composition of the Tsallis entropy 
are described by the, re-scaled, constant negative sectional curvature $k=-1$ space \ $\mathbb{H}^n$. \ It may be of interest to ask how 
generic such a property might be. In particular, what happens if 
we wish to consider the induced metric structure from the Tsallis entropy, not just on 
\ $\mathbb{R}^n$ \ but on some other manifold? The answer is that, the previous analysis remains essentially unchanged.  
This is easily seen, locally at a first,  if we remember that any manifold is locally isometric, to first order, to a Euclidean space [10]. 
The same applies therefore to \ $\mathbb{H}^n$ \ 
which is the constant negative curvature \ $k=-1$ \ manifold, that we reached following the above group-theoretical construction. 
If we wish to consider any other manifold \ $M$ \ with 
constant negative curvature \  $-1$, \ then it will be locally isometric, up to re-scaling, to \ $\mathbb{H}^n$. \  The global structure of \ $M$ \ may 
be quite different though from that of \ $\mathbb{H}^n$, \  the latter being contractible.. 
Any such \ $M$ \  would differ from \ $\mathbb{H}^n$ \ through its non-trivial global topological properties. These non-trivial properties 
are encoded in the fundamental group \ $\pi_1(M)$ \ only, since \ $\pi_i(M)=0, \  i\geq2$ \ as a result of the Cartan-Hadamard theorem [10]. 
Moreover, since \ $M$ \ has same local metric structure as \ $\mathbb{H}^n$, \ its fundamental group should be a subgroup 
of the isometry group of the hyperbolic space \ $\mathbb{H}^n$. \ Since the isometry group of \ $\mathbb{H}^n$ \ is the non-compact Lie group 
\ $SO(n,1)$, \  as can be easily seen by employing 
the hyperboloid model of  \ $\mathbb{H}^n$, \  we conclude that any such \ $M$ \   
must have \ $\pi_1(M)\subset SO(n,1)$. \ Moreover, and in order to avoid orbifold-type singularities on \ $M = \mathbb{H}^n/ \pi_1(M)$ \ the action of 
\  $\pi_1(M)$ \ on \ $\mathbb{H}^n$ \ should be free and properly discontinuous. Determining which exactly subgroups of \ $SO(n,1)$ \ can be 
fundamental groups of \ $M$ \ is a  highly non-trivial issue, but since it is not particularly pertinent for our purposes, 
we will not attempt to foray further toward its answer. It is sufficient for our purposes  to notice that the above construction is universal, 
within the class of Riemannian manifolds induced by (5) and this is essentially guaranteed by the 
Cartan-Hadamard theorem [10]. \\

One immediate consequence of the above construction is that it makes more precise the often-quoted claim that the Tsallis entropy 
describes phenomena that are represented by thick-tailed (temporal and/or spatial) probability distributions [2]. This can be contrasted with the 
BGS entropy which describes very effectively phenomena having ``thin" corresponding tails. Alternatively, the  correlation functions in phenomena 
described by the Tsallis entropy are conjectured to be decaying polynomially (following a ``power-law"), as opposed to exponentially for the BGS entropy 
case. Using the above metric (35), it is easy to see why this occurs. The sets \ $\mathbb{R}$ \ and \ $\mathbb{R}_q$ \ with the 
corresponding operations are field isomorphic and isometric, therefore they are metrically indistinguishable. Consider a phenomenon on 
\ $\mathbb{R}_q^n$ \ having polynomially decaying correlations, with respect to its Euclidean metric. Such a phenomenon is conjecturally 
described by the Tsallis entropy.  The same phenomenon when described 
from the viewpoint of  \ $\mathbb{R}^n$ \ with the  hyperbolic metric (35) has short-range correlations as seen explicitly in (36), 
which shrinks exponentially the distances on \ $\mathbb{R}^n$. \
In turn, this reflects the fact that the unit of distance on \ $\mathbb{R}^n_q$ \ is exponentially larger, by a factor depending on \ $q$ \ via (36)
compared to the one on \ $\mathbb{R}^n$, \ as is also obvious in the definition (7). So the algebraic similarity of the Tsallis entropy to the BGS entropy 
seen when the former is expressed as the q-deformed logarithm
\begin{equation}
   S_q = k_B \ \langle  \ln_q \left( \frac{1}{p_i} \right) \rangle
\end{equation}
where \ $k_B$ \ is the Boltzmann constant and 
\begin{equation}
       \ln_q (x) = \frac{1- x^{1-q}}{q-1}
\end{equation}
is not only an algebraic similarity but it also reflects the common geometric structures induced by the Tsallis and the BGS entropies 
as realized through the field isomorphism (7). \\

Another question that the above construction addresses is how general is really the Tsallis entropy? After all, there are numerous other entropic forms 
that have been used in Information Theory and Physics which share some of the properties of the Tsallis entropy [2] (and references therein). 
Is the Tsallis entropy unique, and if so in which sense? An answer is provided by the Abe [19] and the Santos [20] axioms which play the same role for the 
Tsallis entropy as  the axioms of Khintchin [21] and Shannon [22] respectively for the BGS entropy. Among the assumptions of these sets of axioms, we only 
address the definition of  independence, or composition axiom, which lead to the above construction. 
We see from the above results and the Cartan-Hadamard theorem that the Tsallis entropy should be fairly 
unique. Indeed, its composition (6) gives rise, locally, to the Riemannian metric (35) of  constant negative sectional curvature $k=-1$, up to re-scaling, the 
latter being the universal covering manifold of all possible \ $M$ \ according to the Cartan-Hadamard theorem. \\ 

                                                                                                        \vspace{5mm}


                                      \centerline{\large\sc 4. \   Interactions,  \ Convexity \ and \ CAT($k$) \ Spaces }

                                                                                                           \vspace{5mm}

Continuing the examination of the metric properties induced by (7),  it may be worthwhile to be a bit more general
than we have been so far. There are several reasons for doing so. One is flexibility: we always seek a formalism that is adaptable to 
more general situations than the ones already encountered in order to accommodate future developments. A second reason is that, so far, we 
have treated the case in which a system is described by the Tsallis entropy having a fixed value of \ $q$. \ A natural question that has arisen is what 
happens when two systems that are described by Tsallis entropies having different values of the entropic parameter \ $q$ \ interact with each other [23]. 
More specifically,  is there a value of the entropic parameter \ $q$ \  that describes the combined system, and if so, how would such a value be determined 
[23]? A third reason is the generalization of the previous one for continuous sets of values of \ $q$, \  as in the case of  superstatistics 
[24], [25] , [26] for instance. Generalizing (36), a 
simple two-parameter family of Riemannian metrics induced by  two weakly interacting systems described by Tsallis entropies with non-extensive 
parameters \ $q_1$ \ and \ $q_2$ \  is
\begin{equation}       
  ds^2 = dx^2 + e^{-2t_1x}dy^2 + e^{-2t_2x}dz^2
\end{equation}
where
\begin{equation}
  t_i = \log (2-q_i), \ \ \ i=1,2
\end{equation}
Of course for two such weakly interacting systems, the meaning  of thermodynamic equilibrium, let alone of the rate at which 
such an equilibrium is reached, becomes non-trivial. For a more general class of interacting systems, the corresponding 
induced metric is expected to have a more general form than the doubly warped product (50).  A slightly more general  
Riemannian metric, still of warped-product form, is
\begin{equation}
      ds^2 = dx^2 + h^2_1(x) dy^2 + h^2_2(x,y) dz^2
\end{equation}
where \ $h_1: \mathbb{R}\rightarrow\mathbb{R}$ \ and \ $h_2: \mathbb{R}^2 \rightarrow\mathbb{R}$ \ are convex functions of their arguments.
There are at least three reasons for assuming that \ $h_1$ \ and \ $h_2$ \ are convex. First, convex  functions  have played a
central r\^{o}le in Statistical Mechanics and Thermodynamics, in particular in their axiomatization through properties of the entropy functional [2], [19]-[22], 
 [27], [28], in the conjectural approach to phase transitions via topology changes of the configuration space [29], or in the description of phase 
transitions for systems with few degrees of freedom [30], not to mention  the validity of the Legendre transform and the subsequent (in-)equivalence of 
the classical equilibrium ensembles [31], [32] following the ideas of Gibbs [33]. 
There are several other occasions where the importance of  convexity (or concavity) of the entropy is stressed, but the afore-mentioned 
cases are an indication that convexity is a property of functions of great interest in Statistical Mechanics and Thermodynamics. 
Second, it is worth noticing that in  the metric tensor (35), the warping function is 
negative exponential, so it is convex. Therefore, considering convex functions is a simple and straightforward  extension of (35) and is not entirely 
unreasonable,  although we cannot claim it to be an optimal, let alone a unique, choice. A third reason is simplicity: (continuous) convex functions 
have a unique global minimum, among their other characteristics,  and this property makes them so useful in most of the cases where they are employed in 
Statistical Mechanics and Thermodynamics, as long as we are willing to exclude discussing phase transitions.  \\

Following these considerations, (52) is a doubly warped product on
\begin{equation}
 M = (\mathbb{R} \times_{h_1} \mathbb{R}) \times_{h_2} \mathbb{R}
\end{equation}
 As was explicitly pointed out in the previous paragraph, and can be proved much more generally, for any strictly  convex 
 function \ $h_1(x)$ \ the metric 
 \begin{equation}
   ds^2 = dx^2 + h^2_1(x) dy^2  
 \end{equation}
 has  negative sectional curvature. Because \ $h_2(x,y)$ \ is also assumed to be strictly convex,  according to a theorem of [15], 
 the resulting doubly warped product metric (52) will have negative sectional curvature \ $k_M < 0$. \ Following these arguments, (52) leads us to consider  
 Riemannian manifolds of general, variable, negative sectional curvature. This is still not a sufficiently general 
 construction though. We want to be able consider the thermodynamic, or some other, limit of interest and still get a member of the geometric family under 
 consideration. It is quite simple to see that the limit of a sequence of manifolds with re-scaled metrics  
 is not in general a manifold. Consider, for instance, a flat surface having some localized curvature in the neighborhood of a point [34]. 
 As this curvature converges to a delta function at the point of interest, the corresponding surface develops a metric singularity, thus becoming a flat cone. 
 More generally,  an appropriately defined limit of a sequence of manifolds,
 can develop point-like or even more complicated singularities. A much more robust structure, which is closed under the operation of taking limits
 of sequences, is that of geodesic spaces, namely of length spaces any two points of which are connected by at least one geodesic. 
 Following the previous comments and being motivated by (52), we are interested in geodesic spaces that are negatively curved, 
 or to be more specific, having a uniform negative upper bound on 
 their curvature. There is a very rich theory of such spaces [35], [36] which are called \ CAT(k) \ spaces. They are 
 generalizations of Riemannian manifolds with several desired properties, closure under appropriately defined limits being one of them.\\  

There are various, essentially equivalent, ways to introduce the concept of curvature in (proper) geodesic metric spaces [35], [34]. All are essentially 
abstractions of results in Riemannian geometry which are raised to the status of axioms/definitions whose consequences one seeks to explore. 
Here we only state the triangle comparison condition, 
as we find it to be the most straightforward to formulate and probably the easiest to visualize. 
For alternative definitions and conditions for their equivalence, see [35], [36]. 
Consider a geodesic metric space \ $(X, d_X)$ \ and four points \ $x,y,z,w \in X$. \ Form 
the triangle \ $\triangle xyz$ \ having vertices \ $x, y, z$ \ and as sides the geodesic segments \ $[xy], [yz], [zx]$ \ joining them. Notice that there is no 
requirement for these segments to be unique. Consider a simply-connected Riemannian 
manifold \ $M$ \ with corresponding distance function \ $d_M$ \ having constant sectional curvature \ $k\in\mathbb{R}$, 
\ i.e. a space form, and four points \ $\tilde{x}, \tilde{y}, \tilde{z}, \tilde{w} \in M$. \  Form the 
comparison triangle \ $\triangle \tilde{x}\tilde{y}\tilde{z}$ \ in \ $M$ \ such that \ $d_M(\tilde{x}, \tilde{y}) = d_X(x, y), \  
d_M(\tilde{y}, \tilde{z}) = d_X(y,z), \ d_M(\tilde{z}, \tilde{x}) = d_X(z,x)$ \ having sides the corresponding geodesic segments in \ $M$. \ 
Then \ $(X, d_X)$ \ has curvature less than or equal to \ $k$ \ if \ $\forall \ w \in [yz]$ \ and \ $\tilde{w} \in [\tilde{y}\tilde{z}]$ \ such that \ $d_X(y,w) = 
d_M(\tilde{y}, \tilde{w})$, \ it satisfies \ $d_X(x,w) \leq d_M(\tilde{x}, \tilde{w})$. \ Since we are only interested in negatively curved spaces \ $k<0$
\ no further requirements need be imposed upon \ $(X, d_X)$. \ In short, \ $(X, d_X)$ \ has curvature less than or equal to \ $k<0$ \ if all
geodesic triangles in \ $X$ \ are thinner than the corresponding geodesic triangles of equal sides, one-to-one, in \ $M$.\  Geodesic 
spaces \ $(X, d_X)$ \ for which every point has a neighborhood of curvature \ $\leq k$ \ are called \ CAT(k) \ spaces, from the initials of E. Cartan, A.D. 
Alexandrov and V.A. Toponogov [35], [36].  A common example of CAT(k) spaces are Riemannian manifolds 
of curvature \ $k$. \ Pictorially this is almost obvious, but formally justifying it takes some effort [35]. The CAT(k) definition is non-trivial because it allows  
for other geodesic spaces that do not look like manifolds at all, to fall in this category. One such example of great interest in Geometry and 
Theoretical Computer Science and of considerable potential for Statistical Mechanics, especially for lattice models, are $\mathbb{R}$-trees [35], [36]. 
 These are metric spaces for which\\
 i) there is a unique segment joining each pair of points, and \\
 ii) if two segments have a common point, then their union is also a segment.\\ 
Real trees turn out to be \ CAT(k) \ spaces  for any value of  k, so one could call them CAT($-\infty$) spaces. 
 The converse is also true: a \ CAT(k) \  space for any \ $k\in\mathbb{R}$ \ is an $\mathbb{R}$-tree.
A standard example of spaces that are not \ CAT(k), \ are the Banach spaces of infinite p-summable sequences, 
denoted \ $l^p$, \ with \  $p\geq1$ \ so as to satisfy the  triangle inequality. These turn out not to be \ CAT(k) \ spaces except when \ $p=2$. \ [35].
So we see that the \ CAT(k) \ condition with $\mathrm{k<0}$ is broad enough to include classes of geodesic spaces of considerable interest and still 
specific enough to reflect the metric consequences of (7).    \\         

An obvious question arising at this point is: why should we consider \ CAT(k) \ spaces without demanding a lower curvature bound? 
After all, judging from the Riemannian construction above, there seem to always be a uniform lower bound in the sectional curvature $k$. 
There are several responses to this. First, we observe that 
when we have interacting systems described by the Tsallis entropy with different values of \ $q$ \ the curvature of the semi-direct product metric 
decreases
(i.e. it becomes more negative). This is not too difficult to understand when we observe that the arguments and  the curvature calculations of semi-
direct 
products are special cases of submersions which are well-known [18] to monotonically increase the sectional curvature of the base $B$. Following
the curvature formulae [18] of submersions and the sign conventions in the corresponding Riemannian metric for $q$, we see  
that the curvature of the induced Riemannian metrics should decrease. 
Since there is no a priori upper bound at the value of $q$, although we are examining the case of 
$q\in [0,1)$ in the present work, we have to allow the formalism the flexibility of accommodating \ $q\rightarrow\infty$. \ We would reach the same 
conclusion, 
if we wanted the formalism to be able to accommodate the extreme case of  infinitely many systems interacting with each other,
 each being described by a different value of \ $q$. \ A variation of this theme is the 
case of superstatistics [24], [25], [26], in which the entropic parameter is a function of an externally slowly varying intensive parameter (``temperature") \ $\beta$.  \ 
This 
parameter may stochastically vary following  a \ $\chi^2$ \  density \ $f(\beta )$, \ namely 
\begin{equation}
 f(\beta ) = \frac{1}{\Gamma \left( \frac{1}{q-1} \right)} \left\{ \frac{1}{(q-1)\beta_0} \right\}^\frac{1}{q-1} \beta^{- \frac{2-q}{1-q}}e^{-\frac{\beta}
 {(q-1)\beta_0}}
\end{equation}
where \ $\beta_0$ \ and \ $q>1$ \ are parameters. The range of \ $q>1$ \ does not contradict our analysis of the Tsallis entropy which assumes 
\ $q\in [0,1)$ \ as (55) is only indicative of a possibility, but will not be used in any way in the present work. The Laplace transform of 
the  \ $f(\beta )$ \ stochastically fluctuating temperature, gives rise to the $q$-exponential equilibrium distribution 
\begin{equation}
   \int_0^{\infty} f(\beta ) e^{-\beta E} d\beta = \left\{ 1+(q-1)\beta_0E \right\}^\frac{1}{1-q}  
\end{equation}
This is the $q$-canonical distribution which can be obtained by applying the maximum entropy principle to the Tsallis entropy, 
subject to macroscopic constraints,  and mirrors the derivation of the canonical ensemble in the BGS case [2]. Therefore, considering the Tsallis 
entropy 
induced metric (47) as a special case of superstatistics, we can see the advantages of not imposing a lower curvature bound on the \ CAT(k) \  
spaces under consideration. A second reason for considering \ CAT(k) \ spaces without a lower bound in curvature is that they can be substantially 
different 
from Riemannian spaces but still capture several of the features of negative curvature. Real trees are such an example.  
By contrast, if we consider a \ CAT(k) \ space with both upper and lower 
uniform bounds of its curvature, then  results of [37], [38]  state that outside the singular points, such a space admits a Riemannian structure with 
continuous first derivatives, i.e. of class $C^0$, which coincides with original metric structure on the \ CAT(k) \ space. As a result, if one is willing 
to disregard the lack of regularity associated with the mere continuity of first derivatives (as opposed to at least demanding $C^3$ regularity as is 
usually 
done in Riemannian geometry), then the non-singular part of \ CAT(k) \ spaces with an additional 
uniform lower bound in curvature does not have any major novel geometric features that are not already 
encountered in the Riemannian category. Hence demanding a lower curvature bound in the \ CAT(k) \ spaces is unnecessarily restrictive. 
Third, \ CAT(k) \ spaces are closed under the operation of taking limits. By contrast, manifolds are not. By ``taking limits" we refer to the (pointed) 
Gromov-Hausdorff convergence [36], which is quite appropriate for (proper) compact 
spaces.  More generally, ultralimits with respect to non-principal ultrafilters of  \ $\mathbb{N}$, \ can be used whenever (pointed) Gromov-Hausdorff 
limits 
may not  be defined [35]. As the concepts of (pointed) Gromov-Hausdorff distance/convergence and ultralimits are not explicitly needed in the present 
work, 
we forego their definitions and properties and refer to [35], [36], [39] instead.  It is desirable for such limit spaces to be members of the class of spaces 
under 
consideration, as they may arise as thermodynamic limits of the system under study, or more generally even as limits of the underlying statistical 
system when one or more of  their  parameters approaches value(s) where singularities may arise. There is also a remarkable payback at 
confining ourselves to \ CAT(k) \ spaces: the  concept of an angle can be consistently defined [35], [36] in such spaces. 
As a result, \ CAT(k) \ spaces are not too ``exotic", hence some definitions and results of the familiar geometric and analytic concepts of Euclidean 
spaces still apply to them. To gain some perspective, one may wish to contrast \ CAT(k) \ spaces to other classes of spaces that have recently 
attracted attention such as  Q-L\"{o}wner spaces or more exotic 
examples such as  Bourdon-Pajot buildings or Laakso spaces. Such spaces have some geometrically and analytically familiar properties like 
admitting 
Poincar\'{e} inequalities, or expressions familiar from first order calculus, 
without necessarily resembling Euclidean spaces, as they are far more ``irregular" or ``fractal", by comparison. \\

In closing, we would like to revisit the motivating analogy with Special Relativity presented in Section 2. What we have done  is that we have 
argued that the \ CAT(k), \ $\mathrm{k<0}$ \ spaces should be considered the Tsallis entropy induced metric analogue of the curved manifolds of 
General 
Relativity. Such \ CAT(k) \ spaces have enough flexibility and generality to encode the important metric properties of \ $\tau_q$ \ stemming from (6).
Evidently, there is no point in determining the analogue of Einstein's equations in the above construction, because we developed the relativistic 
analogy just as a metric motivation for reaching (35), without pretending it to have any further, let alone more fundamental, physical significance. \\

                                                                                                           \vspace{5mm}

                                                         
                                                       \centerline{\large\sc 5. \  Discussion \  and \  Further \  Developments }

                                                                                                            \vspace{5mm}

In this work, we constructed the left-invariant  line element (47) initially on \ $\mathbb{R}^2$, \ and then extended it to  \ $\mathbb{R}^n$, \ 
whose properties are encode those of the field isomorphism (7). The corresponding metric turned out to have a constant negative sectional 
curvature depending on \ $q$, \ thus provided a geometric interpretation of the entropic parameter. A further generalization, along the same lines,  
to the case of interacting systems not undergoing phase transitions, characterized by different values of \ $q$, \ lead to manifolds of 
variable sectional curvature. A subsequent generalization 
to a class of spaces locally determined by their curvature, and also closed under appropriate limits has lead us to the \ CAT(k) \ spaces as the category 
of spaces of interest in exploring the metric aspects induced by the Tsallis entropy composition law  (5). 
We have indicated that such spaces play for the metric properties of Tsallis entropy  the same role as the standard Euclidean metric plays for the 
BGS entropy. Thus, the Cartan-Hadamard theorem accounts for the universality and uniqueness (in a sense) of Tsallis' entropy, as is encoded 
in the Abe [19] and Santos [20] axioms.\\

One point that has to be addressed is to what extent the metric properties of (7) examined in the present work are particular to Riemannian metrics. 
More concretely, if we insisted in using (12) instead of its Riemannian counterparts (47), (52), how many of the conclusions of 
the current work could be carried through? The answer seems to be: essentially all of them, except any regularity statements directly tied to the 
Riemannian metric. This is due to the fact that the properties examined in the present work depend on a far more general and robust structure than 
negative curvature:  hyperbolicity. The details of how this comes about, as well as a re-assessment of the geometric reasons and the 
general framework for the uniqueness of the Tsallis entropy will be presented in a future work. As a by-product 
of this approach, will also be able to naturally address the issue of the 
positive-definiteness of the metric (37) and its behavior far from the origin as was noticed in the discussion following (37). 
Based on the hyperbolicity, will also be able to relate the 
metric properties with the doubling measure properties induced by \ $\tau_q$ \ in \ CAT(k) \ spaces, continuing the preliminary analysis that took place in
[5] in the case of  \  $\mathbb{R}$.\\
   
                                                                                                           \vspace{10mm}
  
  
                                                                                     \centerline{\large\sc Acknowledgements}
  
                                                                                                          \vspace{3mm}
  
  We would like to express our gratitude to Professor Luiz C. L. Botelho for correspondence after the completion of the present work, 
  that brought to our attention ref. [26]. Based on the time of writing and its content, we believe that [26]  
  should be counted among the fundamental papers that eventually lead to the concept of superstatistics.\\ 
  
                                                                                                         \vspace{5mm}
  

                                                                                                         \newpage

                                                                                         \centerline{\large\sc References}
      
                                                                                                        \vspace{3mm}

\noindent [1] \  C. Tsallis, \  J. Stat. Phys. {\bf 52}, \ 479 \  (1988)  \\
\noindent [2] \  C. Tsallis, \ \emph{Introduction to Nonextensive Statistical Mechanics:  Approaching a Complex\\ 
                              \hspace*{5mm} World}, \ Springer (2009)\\
\noindent [3] \ C. Tsallis, \ \emph{Some Open Points In Nonextensive Statistical Mechanics}, \ arXiv:1102.2408v1\\                                                  
\noindent [4] \  T.C. Petit Lob\~{a}o, P.G.S. Cardoso, S.T.R. Pinho, E.P. Borges, \ Braz. J. Phys. {\bf 39}, \\
                               \hspace*{6mm} 402 (2009)\\
\noindent [5] \ N. Kalogeropoulos, \ Physica A {\bf 391}, \ 1120 \ (2012)  \\
\noindent [6] \ L. Nivanen, A. Le M\'{e}haut\'{e}, Q.A. Wang, \ Rep. Math. Phys. {\bf 52}, 437 (2003)\\
\noindent [7] \ E.P. Borges, \ Physica A {\bf 340}, 95 (2004)\\ 
\noindent [8] \ N. Kalogeropoulos, \ Physica A {\bf 356}, 408 (2005)\\
\noindent [9] \ P.G.S. Cardoso, T.C.Petit Lob\~{a}o, E.P. Borges, S.T.R. Pinho,  \ J. Math. Phys. {\bf 49}, \\
                                  \hspace*{6mm} 093509 (2008)\\   
\noindent [10] \ J. Cheeger, D.G. Ebin, \ \emph{Comparison Theorems in Riemannian Geometry}, \\ 
                                   \hspace*{8mm} Amer. Math. Soc. Chelsea Publ.  (1975)\\
\noindent [11] \ A.O. Barut, R. Raczka, \ \emph{Theory of Group Representations and Applications}, \\
                                  \hspace*{8mm} 2nd Ed., \  World Scientific (1986)\\
\noindent [12] \  R. Azencott, E.N.  Wilson, \ Trans. Amer. Math. Soc. {\bf 215}, 323 (1976)\\ 
\noindent [13] \  R. Azencott, E.N.  Wilson, \ \emph{Homogenous manifolds with negative curvature, \\
                                 \hspace*{8mm} Part II}, \  Mem. Amer. Math. Soc.  {\bf 8},  (1976)\\
\noindent [14] \ E. Heintze, \  Math. Ann. {\bf 211}, 23 (1974)\\
\noindent [15] \ R.L. Bishop, B. O'Neill, \ Trans. Amer. Math. Soc. {\bf 145}, 1  (1969)\\ 
\noindent [16] \ H. Stephani, D. Kramer, M. MacCallum, C. Hoenselaers, E. Herlt, \ \emph{Exact Solutions\\
                                 \hspace*{8mm} of Einstein's Field Equations}, 2nd Ed., \ Cambridge Univ. Press (2003)\\
\noindent [17] \  K. Becker, M. Becker, J. Schwarz, \ \emph{String Theory and M-Theory}, \\
                                  \hspace*{8mm} Cambridge Univ. Press (2007)\\
\noindent [18] \  B. O'Neill, \ Michigan Math. J. {\bf 13}, 459 (1966)\\ 
\noindent [19] \ S. Abe, \ Phys. Lett. A {\bf 271}, 74 (2000)\\
\noindent [20] \ R.J.V. Santos, \ J. Math. Phys. {\bf 38}, 4104 (1997)\\ 
\noindent [21]  \ A.J. Khintchin, \ Usp. Mat. Nauk. {\bf 8}, 3 (1953)\\
\noindent [22] \ C.E. Shannon, \ Bell Syst. Tech. J.  {\bf 27}, 379 (1948); \ \ Ibidem {\bf 27}, 623 (1948)\\
\noindent [23] \ G.-A. Tsekouras, C. Tsallis, \ Phys. Rev. E {\bf 71}, 046144 (2005)\\
\noindent [24] \ C. Beck, E.G.D. Cohen, \ Physica A {\bf 322}, 267 (2003) \\
\noindent [25] \ C. Beck, \ \emph{Recent developments in superstatistics}, \ arXiv:0811.4363\\
\noindent [26] \ L.C.L. Botelho, \ Mod. Phys. Lett. B {\bf 17}, 733 (2003)\\ 
\noindent [27] \ E. H. Lieb, J. Yngvason, \ Phys. Rep. {\bf 310}, 1 (1999)\\
\noindent [28] \ G. Ruppeiner, \ Amer. J. Phys. {\bf 78}, 1170 (2010)\\
\noindent [29] \ M. Kastner, \ Rev. Mod. Phys. {\bf 80}, 167 (2008)\\
\noindent [30] \ D.H.E. Gross, \ Entropy {\bf 6}, 158 (2004)\\
\noindent [31] \ A. Campa, T. Dauxois, S. Ruffo, \ Phys. Rep. {\bf 480}, 57 (2009)\\
\noindent [32] \ H. Touchette, \  \emph{Ensemble equivalence for general many-body systems}, \ arXiv:1106.2979\\                     
\noindent [33] \ J.W. Gibbs, \ \emph{Elementary Principles in Statistical Mechanics},  Yale Univ. Press (1948) \\  
\noindent [34] \ V.N. Berestovskij, I.G. Nikolaev, \ \emph{Multidimensional Generalized Riemannian Spaces},\\  
                    \hspace*{8mm}  in  \emph{Geometry IV: Nonregular Riemannian Geometry}, Y.G. Reshetnyak (Ed.), Encycl.\\
                    \hspace*{8mm}  Math. Sci., \  Vol. {\bf 70},  \ Springer \ (1993)\\             
\noindent [35] \ M. R. Bridson, A. Haefliger, \ \emph{Metric Spaces of Non-Positive Curvature}, Springer (1999)\\
\noindent [36] \  M. Gromov, \ \emph{Metric Structures for Riemannian and Non-Riemannian Spaces},\\
                     \hspace*{8mm}  Birkh\"{a}user \ (1999)\\    
\noindent [37] \ Y. Otsu, T. Shioya, \ J. Diff. Geom. {\bf 39}, 629 (1994)\\                     
\noindent [38] \ Y. Otsu, \  Math. Sci. Res. Inst. Publ.  {\bf 30},  135 (1997)\\              
\noindent [39] \ K. Fukaya, \  Adv. Studies Pure Math. {\bf 18}, 143 \ (1990)\\

                                        \vfill

\end{document}